\documentclass{article}
\usepackage{arxiv}
\usepackage{nameref}
\usepackage[utf8]{inputenc} %
\usepackage[T1]{fontenc}    %
\usepackage{color,soulutf8}
\definecolor{myred}{RGB}{216, 28, 56}  
\definecolor{myblue}{RGB}{62,73,173}%
\usepackage[colorlinks=false,breaklinks=true,hyperfootnotes=true,allcolors=myblue]{hyperref}      %
\usepackage{url}            %
\usepackage{booktabs}       %
\usepackage{amsfonts}       %
\usepackage{nicefrac}       %
\usepackage{microtype}      %
\usepackage{lipsum}		    %
\usepackage{graphicx}
\usepackage[square,numbers,compress]{natbib}
\usepackage{doi}
\usepackage[inline]{enumitem}
\usepackage{subcaption}
\usepackage[export]{adjustbox}
\usepackage{amsmath}
\usepackage{booktabs}%
\usepackage{cleveref}

\title{Space-based gravitational wave signal detection and extraction with deep neural network}

\date{\today}	%

\author{
    Tianyu Zhao\textsuperscript{1,2,3}\thanks{These authors contributed equally.}, 
    Ruoxi Lyu\textsuperscript{4}\footnotemark[1], 
    He Wang\textsuperscript{5,6,7} 
    Zhoujian Cao\textsuperscript{1,2,8}\thanks{Corresponding author:  \href{mailto:zjcao@amt.ac.cn}{\texttt{zjcao@amt.ac.cn}};},
    Zhixiang Ren\textsuperscript{3}\thanks{Corresponding author: \href{mailto:renzhx@pcl.ac.cn}{\texttt{renzhx@pcl.ac.cn}};}
    \bigskip
    \\ \textsuperscript{1} Department of Astronomy, Beijing Normal University, Beijing, 100875, China 
    \\ \textsuperscript{2} Institute for Frontiers in Astronomy and Astrophysics, Beijing Normal University, Beijing, 102206, China 
    \\ \textsuperscript{3} Peng Cheng Laboratory, Shenzhen, 518055, China
    \\ \textsuperscript{4} Department of Statistics, University of Auckland, Auckland, 1142, New Zealand
    \\ \textsuperscript{5} International Centre for Theoretical Physics Asia-Pacific, 
    \\ University of Chinese Academy of Sciences (UCAS), Beijing, 100190, China
    \\ \textsuperscript{6} Taiji Laboratory for Gravitational Wave Universe, UCAS, Beijing, 100049, China 
    \\ \textsuperscript{7} CAS Key Laboratory of Theoretical Physics, Institute of Theoretical Physics,
    \\ Chinese Academy of Sciences, Beijing, 100190, China
    \\ \textsuperscript{8} School of Fundamental Physics and Mathematical Sciences, Hangzhou Institute for Advanced Study,
    \\ UCAS, Hangzhou, 310024, China
}

\renewcommand{\shorttitle}{Space-based GW signal detection and extraction with DNN}

\begin{document}
\maketitle

\begin{abstract}
    Space-based gravitational wave (GW) detectors will be able to observe signals from sources that are otherwise nearly impossible from current ground-based detection.
    Consequently, the well established signal detection method, matched filtering, will require a complex template bank, leading to a computational cost that is too expensive in practice.
    Here, we develop a high-accuracy GW signal detection and extraction method for all space-based GW sources.
    As a proof of concept, we show that a science-driven and uniform multi-stage self-attention-based deep neural network can identify synthetic signals that are submerged in Gaussian noise.
    Our method exhibits a detection rate exceeding 99\% in identifying signals from various sources, with the signal-to-noise ratio at 50, at a false alarm rate of 1\%. while obtaining at least 95\% similarity compared with target signals.
    We further demonstrate the interpretability and strong generalization behavior for several extended scenarios.
\end{abstract}


\section{Introduction}
\label{sec1}
The first direct detection of GWs coming from coalescing binary black holes (BBHs) was made by the LIGO/Virgo Collaboration \cite{abbott_observation_2016}, which verifies Einstein's General Relativity.
As detectors become more sensitive, more and more GW events are discovered, enabling a new era of multi-messenger astronomy.
A total of 93 events have been reported in the three observations \cite{the_ligo_scientific_collaboration_gwtc-3_2021}. GWs have become a new probe allowing cross-validation with a variety of fundamental physical theories \cite{the_ligo_scientific_collaboration_tests_2016, the_ligo_scientific_collaboration_astrophysical_2016,bailes_gravitational-wave_2021,arunNewHorizonsFundamental2022}.

Ground-based GW detectors such as LIGO, Virgo, and KAGRA cannot detect GWs at frequencies lower than $10\mathrm{Hz}$ due to seismic noise \cite{Matichard_2015}, therefore space-based detectors are being developed.
Laser Interferometer Space Antenna (LISA) will be launched around 2034 \cite{amaro-seoane_laser_2017}, and Taiji \cite{hu_taiji_2017} and TianQin \cite{luo_tianqin_2016} are also in progress.
LISA is expected to observe a variety of GW sources \cite{bambi_space-based_2021}, including Galactic binaries (GB), massive black hole binaries (MBHB), and extreme-mass-ratio-inspirals (EMRI).
The most common GB sources are binary white dwarf (BWD) systems, which will populate the whole frequency band of the LISA detector.
Massive black holes (MBH) exist in most galactic centers, and the MBHs merge along with the galaxies, which happens regularly in the Universe \cite{PhysRevD.93.024003}.
The EMRI system is formed when the MBH captures compact objects (CO) surrounding them.
Unlike stars, COs can avoid tidal disruption and approach the central MBH, radiating a significant amount of energy in GWs at low frequencies.
Beyond these resolvable sources, a huge number of unresolvable events will sum up incoherently, forming a stochastic GW background (SGWB).
The detection of those GWs in the LISA mission enables us to gain a better understanding of black holes and galaxies \cite{pan_formation_2021}.

GW data processing is complicated due to the overwhelming noise, which is non-Gaussian, sometimes non-stationary \cite{zevin_gravity_2017,ormiston_noise_2020}, and containing sudden temporary glitches \cite{zevin_gravity_2017,ormiston_noise_2020,mogushi_reduction_2021}.
Earlier GW detection methods are divided into two categories:
\begin {enumerate*} [label=\itshape\alph*\upshape)]
\item theoretical template-based algorithms like matched filtering,
\item template-independent algorithms.
\end {enumerate*}
In principle, the most accurate results can be achieved by using a matched filtering algorithm to detect signals buried in Gaussian noise \cite{finn_detection_1992,usman_pycbc_2016,cannon_gstlal_2021}.
This is currently the most widely used algorithm for the detection of GWs.
The additional complexity of space-based detection over ground-based detection can be attributed to the different types of sources.
The optimal template for matched filtering would have to include all the GW source parameters in the data.
However, this is not practical because of the high parameter space dimension to be explored.
Moreover, the typical duration of the compact binary coalescence signal detected by LISA is longer than that detected by ground-based detectors, resulting in an even larger computational effort for the matched filtering algorithm.
For ground-based detection, a series of template-independent signal extraction and detection algorithms have been developed, such as CWb \cite{klimenko_coherent_2008}, and BayesWave \cite{cornish_bayeswave_2015}, both based on the wavelet transformation.
Ref. \cite{torres_total-variation-based_2014} proposes a total-variation-based method, and a novel approach based on the Hilbert-Huang transform was recently developed \cite{Akhshi2021}.
The advantage of these algorithms is that they are not limited by theoretical template banks and can extract the signal from noisy data, or, in other words, reconstruct the signal waveform. The disadvantage of these algorithms is that they are only available for burst signals, which are not suitable for space-based GW signals.

Deep learning has already been successful in various GW data analyses, such as signal detection \cite{george_deep_2018-1, gabbard_matching_2018, wang_gravitational_2020,krastev_real-time_2020, lopez_deep_2021, skliris_real-time_2021, zhang_detecting_2022}, parameter estimation \cite{gabbard_bayesian_2022, dax_real-time_2021}, glitch classification \cite{colgan_efficient_2020, cavaglia_finding_2019, razzano_image-based_2018}, noise reduction \cite{ormiston_noise_2020,mogushi_reduction_2021}, and signal extraction \cite{torres-forne_denoising_2016, wei_gravitational_2020, shen_denoising_2019,chatterjee_extraction_2021}.
Several deep neural network techniques have been used in signal extraction, including dictionary learning \cite{torres-forne_denoising_2016}, WaveNet \cite{wei_gravitational_2020}, denoising autoencoder \cite{shen_denoising_2019}, and LSTM \cite{chatterjee_extraction_2021}.
Most of these existing methods for detection\cite{george_deep_2018-1,wang_gravitational_2020,krastev_real-time_2020,lopez_deep_2021} are designed for ground-based GWs, and for space-based signal, \cite{zhang_detecting_2022} and \cite{ruan_rapid_2023} only achieve single source detection of EMRI and MBHB signals, respectively. In Ref. \cite{khan_interpretable_2022}, the transformer-based model is first applied to GW waveform forecasting.
However, there is no unified treatment for all sources of the space-based GW signal detection and extraction method.

In this article, we develop a uniform deep learning-based model for space-based GW signal detection and extraction for the four main GW sources of LISA.
Our model is based on a self-attention sequence-to-sequence architecture that performs well when dealing with time series data. We have integrated long-term and short-term feature extraction blocks in our model to catch the dependency of the GW signal in high-dimensional latent space.
To our knowledge, this is the first study to achieve high-accuracy detection and high-precision extraction for all main potential GW signals from space-based detection.
The model's intermediate results can be interpreted as the encoded signal waveform, revealing a strong correlation between what needs to be learned and what has been learned by the neural network.
In our test results analysis, we obtained average overlaps (see equation \eqref{eq:overlap}) of 95\% of our test samples being greater than $0.95$.
It takes less than $10^{-2}$ seconds to perform extraction and detection, which is a factor of roughly $10^5$ improvement over traditional approaches that often require several hours.
Finally, our method can also achieve considerable extraction effects for some signals generated by other models that are not in the training dataset, demonstrating the strong generalization ability of the model.

\section{Results}
\subsection{Deep neural network}
\begin{figure*}[htbp]
    \centering
    \begin{subfigure}[t]{0.98\textwidth}{
            \includegraphics[width=\textwidth,valign=t]{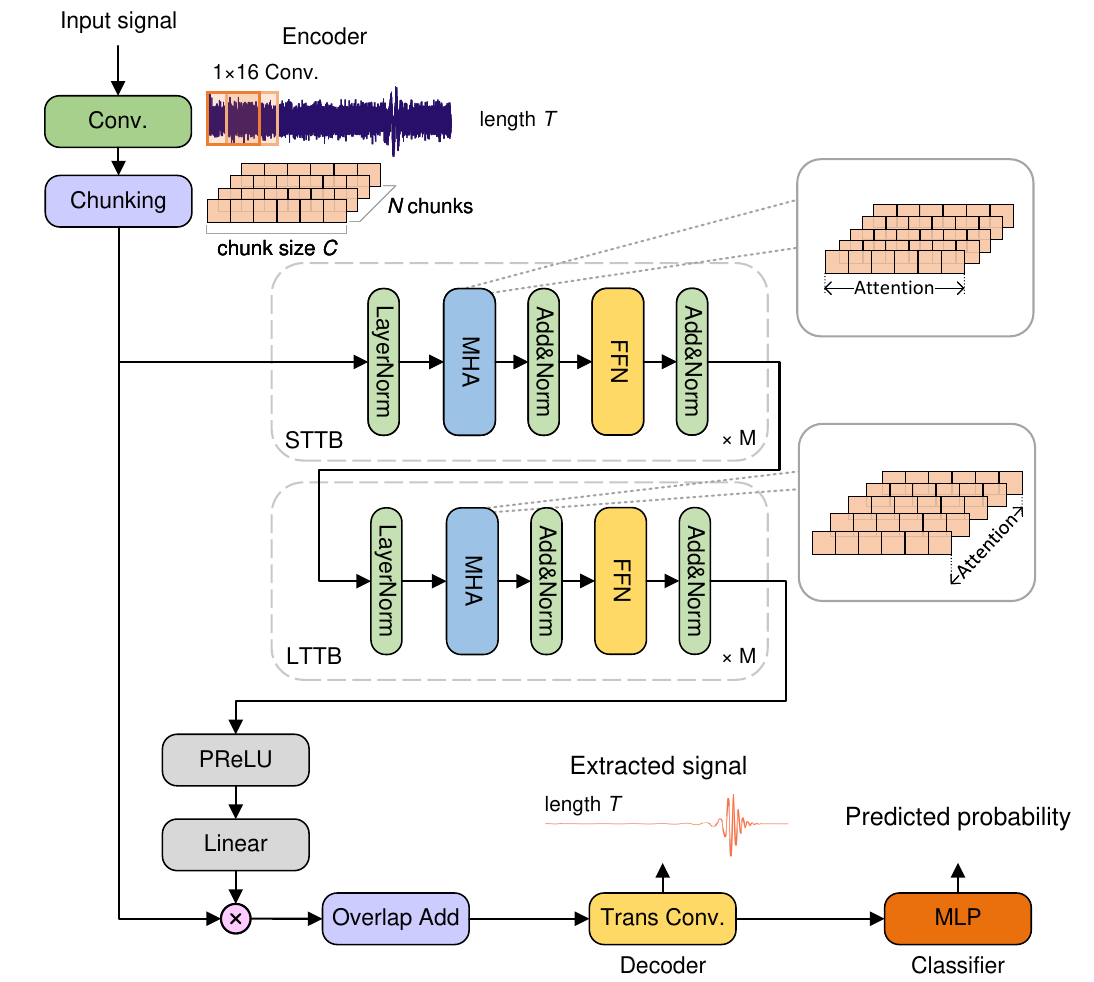}
        }
    \end{subfigure}
    \caption{
        \textbf{The overall architecture of our Transformer based deep neural network.} Beginning with a convolutional network-based encoder, data is transformed and feed into the Transformer-based extraction network. This network composed of several Short-Term Transformer Blocks (STTB) and Long-Term Transformer Blocks (LTTB), excels in capturing both local and global dependencies within the GW data, aimed at extracting GW signals. The final stage is a multi-layer perception-based classifier, responsible for the signal detection and provide a predicted probability.
    }
    \label{fig:network}
\end{figure*}

We extend the mask-based prediction method for speech enhancement in Conv-TasNet\cite{luo2019conv} with a self-attention-based network for our task. As shown in Fig. \ref{fig:network}, the network consists of four processing stages: encoder, extraction net, decoder, and classifier. First, the encoder maps the signal from the detector to a high-dimensional representation and splits it into short chunks. This encoded representation is used to estimate a mask network for signal extraction. The extraction network produces a mask matrix with Transformer blocks catching both short-term and long-term dependency from chunks. The decoder uses a transposed convolution layer for the element-wise multiplying of the mask matrix and an encoded representation to reconstruct the extracted signal. Finally, the extracted signal is sent to a multi-layer linear perception classifier for detection. The classifier gives the predicted probability that the input data contains a true GW signal.

\subsection{Generate space-based GWs dataset}
Due to the large differences in signals of GWs from different sources, we generate training and testing datasets for each source.
We choose a universal sampling rate of $0.1{\rm Hz}$ for all datasets.
All the data samples in the datasets have 16000 sampling points, hence the duration of each sample is 160000 seconds or 44.4 hours.
The parameter space of signal generation is sampled with a uniform grid.
The range of the parameters used to generate the GW signals are listed in the Table. \ref{tab:emri_par}-\ref{tab:bwd_par}.
Then, the parameters on each grid point are used to generate the corresponding GW signals.
For noise generation, we use the noise power spectral density (PSD) of LISA \cite{robson_construction_2018} to simulate Gaussian noise. It should be noted that in this study, galactic confusion noise has not been taken into consideration.
To simulate the different signal-to-noise ratio (SNR) levels in signal extraction, we set the SNR equal to 50, 40, and 30 following the equation \eqref{eq:snr}.

\begin{figure}[htbp]
    \centering
    \begin{subfigure}[t]{0.99\textwidth}{
        \includegraphics[width=\textwidth,valign=t]{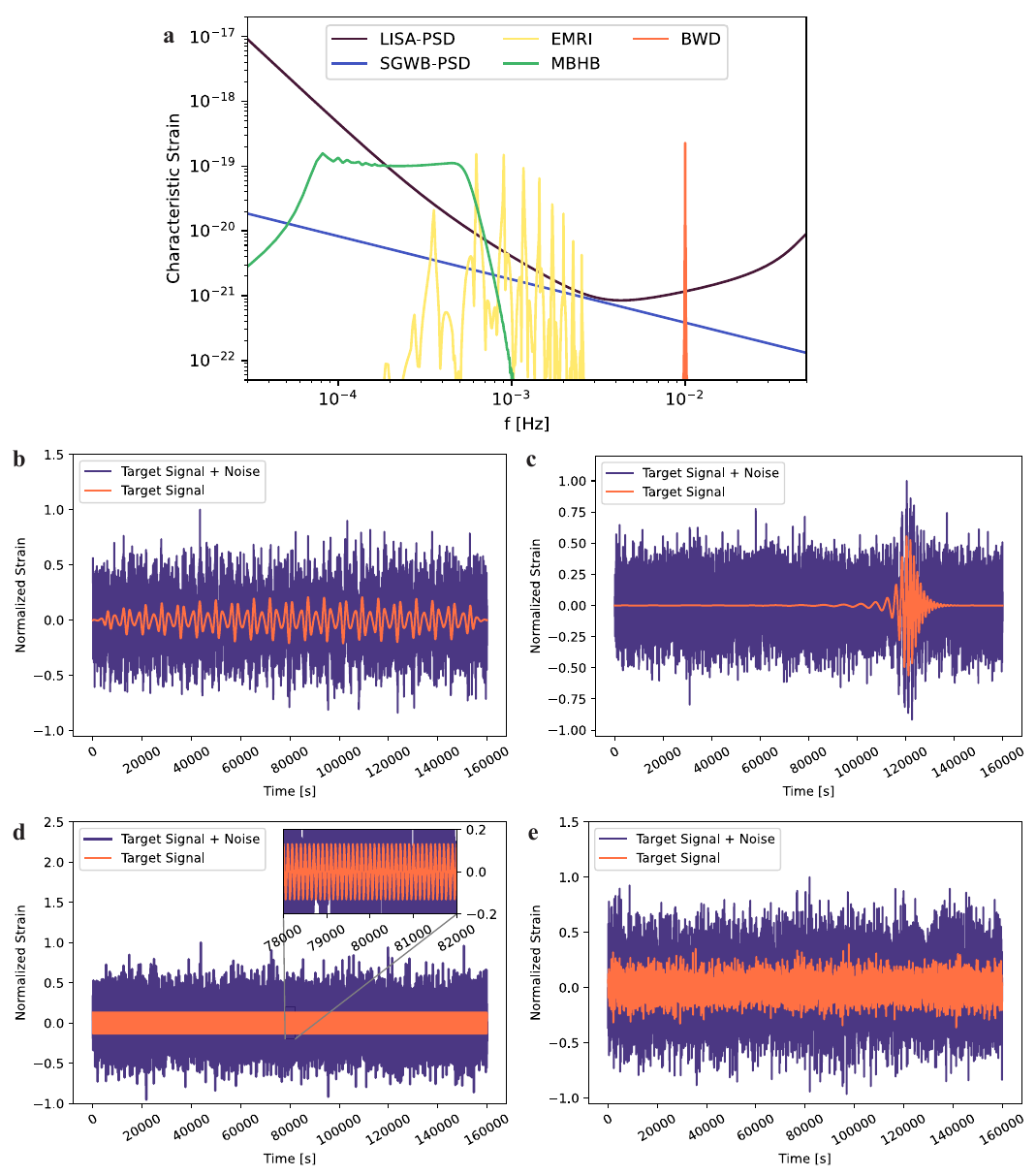}
    }
    \end{subfigure}
    \caption{
        \textbf{Training data samples of each GW source in the frequency domain and time domain.} \textbf{a}, an example of EMRI, MBHB, and BWD signal power spectra along with SGWB power spectral density (PSD) and LISA sensitivity curve in the frequency domain. \textbf{b-e}, each sub-figure shows an example of a GW signal from a specific source. Here we plot the whitened $h_I(t)$ (orange) and $h_I(t)$ with noise (purple). The signal-to-noise ratio (SNR) of EMRI, MBHB, and BWD signals is $50$, and the SGWB signal has $n_t=2/3$ and $\alpha=-11.35$ (see equation \eqref{eq:sgwb_pl}).
    }
    \label{fig:data}
\end{figure}

The SGWB data is directly generated by the PSD given by equation \eqref{eq:sgwb_pl}, in which the parameter $\alpha$ characterizes the amplitude of the SGWB signal. We set $\alpha$ equal to -11.35, -11.55, and -11.75 to generate the signal in the test datasets.
To train the model for the GW detection task, we generate that half of the samples contain signals and half are pure noise. Fig. \ref{fig:data} shows some sample cases for our generated data. Specific details for each dataset will be presented in \nameref{sec:method}.

\subsection{Interpretability of the network}
\begin{figure}[htbp]
    \centering
    \begin{subfigure}[t]{0.99\textwidth}
        \includegraphics[width=\textwidth,valign=t]{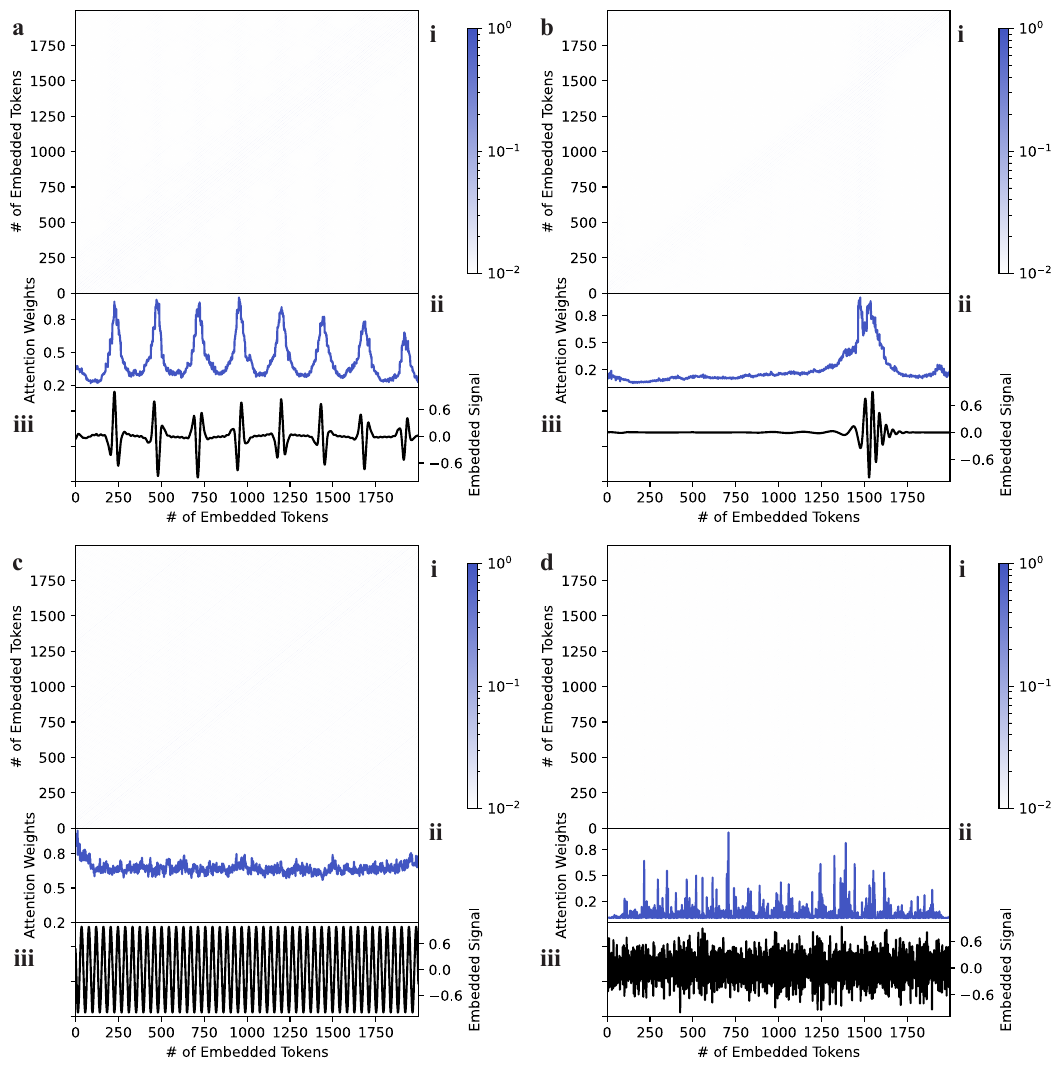}
    \end{subfigure}
    \caption{
        \textbf{The interpretability of the deep neural network.} \textbf{a-d}, each sub-figure demonstrates the intermediate results of a model trained on specific sources, where the corresponding salient features represent the physical signal in the time domain. Each sub-figure consist of 3 sub-sub-panels, shows the attention map (panel \textbf{i}), the average attention received by each token from all tokens (panel \textbf{ii}), and the corresponding embedded signal contained in the input data (panel \textbf{iii}).
    }
    \label{fig:inter-att}
\end{figure}

To better understand what information the self-attention-based neural network learned, we explored the corresponding relationship between the attention mechanism and the embedded input signal. We calculate the attention maps of various layers of the network. The attention map presents the average output of attention heads in each layer, between each pair of tokens. The output of each attention head is a weighted sum of the embedded tokens of signal, which will be defined in detail in the \nameref{sec:method}.

The core of our network is the extraction net, consisting of several STTBs (Short-Time Transformer Block) and LTTBs (Long-Time Transformer Block) stacked together. Both LTTB and STTB contain multi-head attention layers, which enable our model to have universal learning capability for different GW sources. In STTB, attention is only calculated between tokens in the same chunk, which indicates that all tokens exchange information within the corresponding chunk. The attention map of STTBs is a diagonal line consisting of squares in chunk size, implying STTBs are only interested in local structures. And the top panel of each sub-figure in Fig. \ref{fig:inter-att} shows the attention maps between all embedded tokens in LTTB. We can see that for different GW sources, the model weights show different patterns. We sum each column of the attention map to obtain the attention received by each token, which is called attention weights (middle panel). As seen for the EMRI, MBHB, and BWD models, the attention weights and the signal after a sliding average follow the same pattern. We also applied the Augmented Dickey-Fuller (ADF) test to the summed attention map matrix, which is a $1\times1999$ vector, for both the BWD and SGWB models, owing to the stationary nature of the signals in the data. These results were -6.84 and -15.33, respectively, indicating that the attention maps are also stationary. This means that our LTTB can learn the global dependency of the data. The above experiments show that our self-attention-based model has the ability to learn both local and global structures for different physical scenarios.

\subsection{GW signal extraction}
\begin{figure}[htbp]
    \centering
    \begin{subfigure}[t]{0.99\textwidth}{
            \includegraphics[width=\textwidth,valign=t]{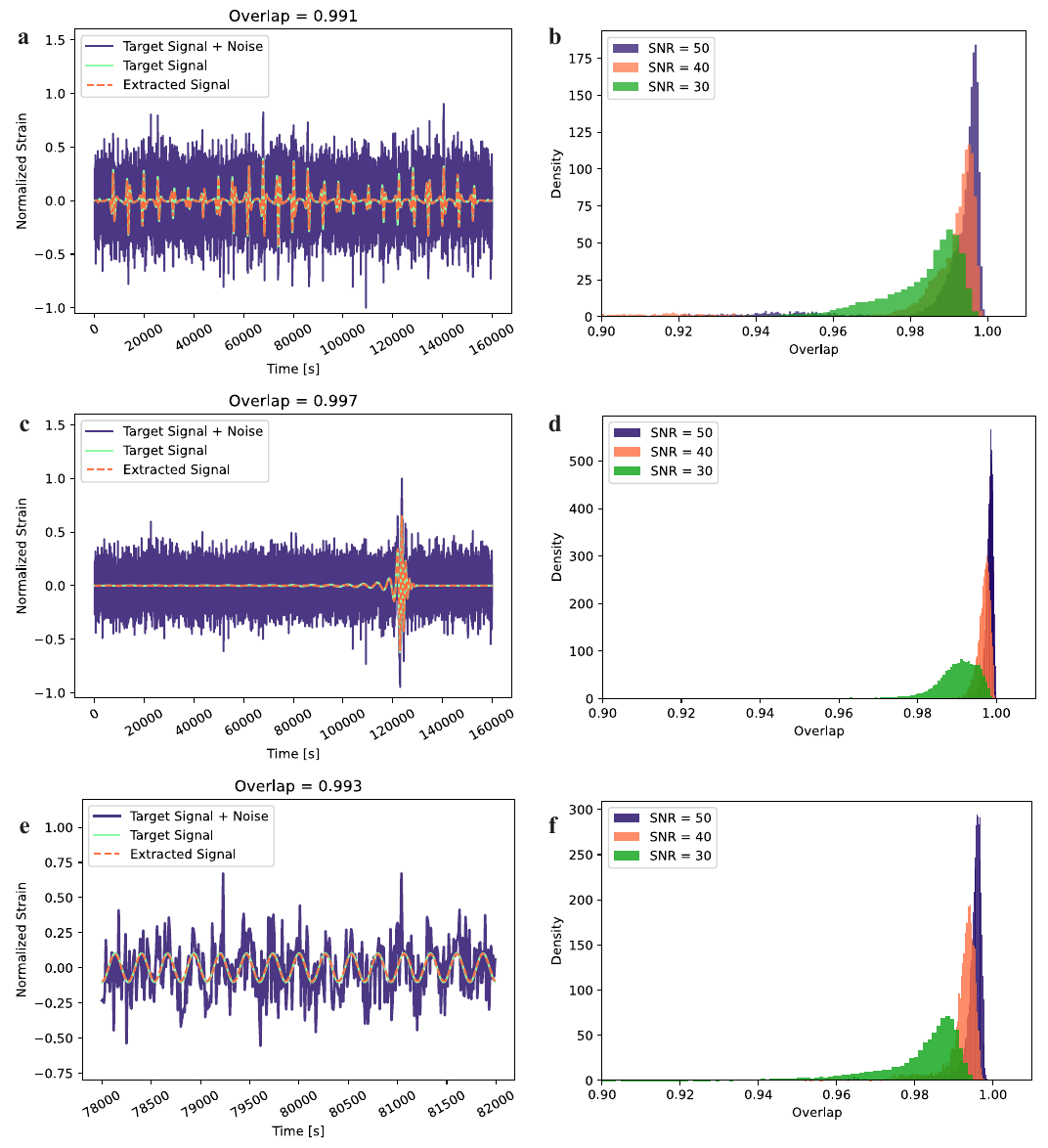}}
    \end{subfigure}

    \caption{
        \textbf{The signal extraction examples and the overlapping (between the target and extracted signals) distributions for different GW sources.} \textbf{a} and \textbf{b}, EMRI. \textbf{c} and \textbf{d}, MBHB. \textbf{e} and \textbf{f}, BWD. The extracted signal is compared with whitened templates. Only the middle part of the BWD waveform is presented to show the details of the waveform. The overlap between extracted data and waveform templates is shown on the top.
        The high values indicate the strong performance of our method on signal extraction for different GW sources. Tests on different signal SNRs also show our models' generalization ability.}
    \label{fig:denoise}
\end{figure}

The most straightforward test is to use our model to extract a piece of data with Gaussian noise and see whether it is able to extract the signal.In Fig. \ref{fig:denoise} we show some examples of the signal extraction effect of our model for different types of GW signals. Three subplots a,b, and c show the signal extraction effect of our model for EMRI, MBHB, and BWD signals, and the overlap between the model output and the template is calculated. The overlap shows a very good signal extraction effect. Since SGWB does not have a specific waveform, we do not do this test here for the corresponding model. The most straightforward test is to use our model to extract a piece of data with Gaussian noise and see whether it is able to extract the signal.In Fig. \ref{fig:denoise} we show some examples of the signal extraction effect of our model for different types of GW signals. Three subplots a,b, and c show the signal extraction effect of our model for EMRI, MBHB, and BWD signals, and the overlap between the model output and the template is calculated. The overlap shows a very good signal extraction effect. Since SGWB does not have a specific waveform, we do not do this test here for the corresponding model.

Then, we perform some statistical tests on the model. We generate the test data in the same way as we generate the training dataset. We use a coarser grid and create test datasets of $10,000$ samples containing signals with SNR equal to 30, 40, and 50 respectively for each type of GW. In the left column of Fig. \ref{fig:denoise}, the signal extraction effect of the three models for each of the three types of GWs is depicted. The showcase is selected according to the 15 percent quantile of the testing overlap of each GW source. In the right column, it can be seen that for the case of ${\rm SNR}=50$, the signal extraction effect is performed with great accuracy for MBHB signals, and the overlap is higher than $0.99$ for all test samples. The overlap for the BWD signal is greater than $0.97$ for $95\%$ of the test samples. Although it is slightly less effective in extracting the EMRI signal due to its complexity, $92\%$ of samples have overlap greater than $0.95$.

\subsection{GW signal detection}
\begin{figure}[htbp]
    \centering
    \begin{subfigure}[t]{0.99\textwidth}{
        \includegraphics[width=\textwidth,valign=t]{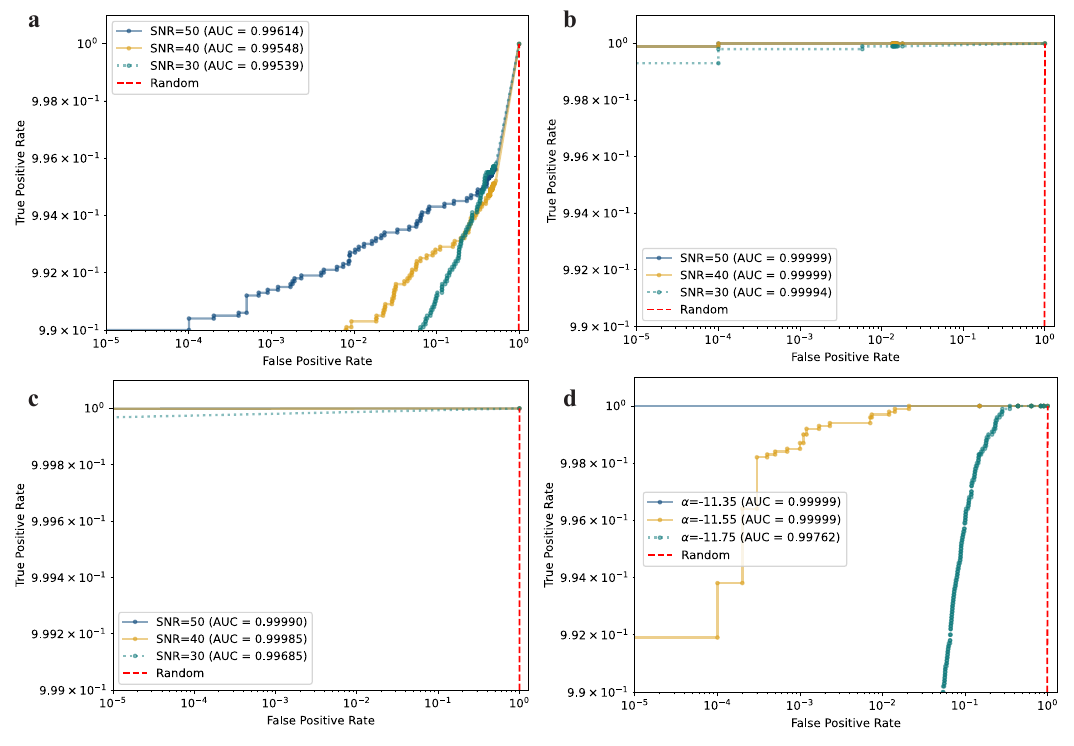}
    }
    \end{subfigure}

    \caption{
        \textbf{The signal detection performance from our classification perspective.} \textbf{a-c}, each sub-figure shows the ROC curve of a model aimed at detecting EMRI, MBHB, and BWD signals with test data SNR equal to 30, 40, and 50 respectively. \textbf{d}, the sub-figure shows the ROC curve of a model aimed at detecting SGWB with 3 different amplitudes.
        We show the ROC curve on a logarithmic scale to better visualize the shape at a lower false alarm rate.
        The high AUC values indicate the strong ability of our method for multiple GW source detection.
    }
    \label{fig:detection}
\end{figure}

The signals extracted by the neural network are used to build up four new datasets to test the ability of our model to detect GW signals. These four test datasets each included 10,000 samples containing signals and 10,000 samples containing pure noise.
Here we utilize the detection rate, another term for true positive rate (TPR), to measure the probability of correctly identifying signals.
For each signal type and an SNR of (30, 40, 50) and a false alarm rate at 1\%, the detection rates are (98.20\%, 99.70\%, 99.71\%) for EMRI, (99.99\%, 100\%, 100\%) for MBHB, (99.37\%, 99.97\%, 99.98\%) for BWD, respectively.
For SGWB signals with $\alpha$ of (-11.75, -11.55, -11.35) and a false alarm rate at 1\%, the detection rates are (95.05\%, 99.97\%, 100\%), respectively.
To quantify the performance of our model in detection tasks, we use the receiver operating characteristic (ROC) curve as shown in Fig. \ref{fig:detection}.
In ROC analysis, the true positive rate (TPR) and the false positive rate (FPR) are plotted as the probability threshold for classifying a candidate as positive (i.e. signal) is altered.
The area under the ROC curve (AUC) has been used to evaluate the classifier's performance, which is a single scalar value between 0 and 1.
In general, the higher the AUC, the better the classifier.
We calculate AUC using \texttt{Scikit-Learn} library \cite{scikit-learn}.
For all 4 signal types, the AUC are all close to $1$, which indicates a fairly high sensitivity for classification (signal detection).

\subsection{Test on LDC2a}
\begin{figure}[htb]
    \centering
    \includegraphics[width=0.95\textwidth]{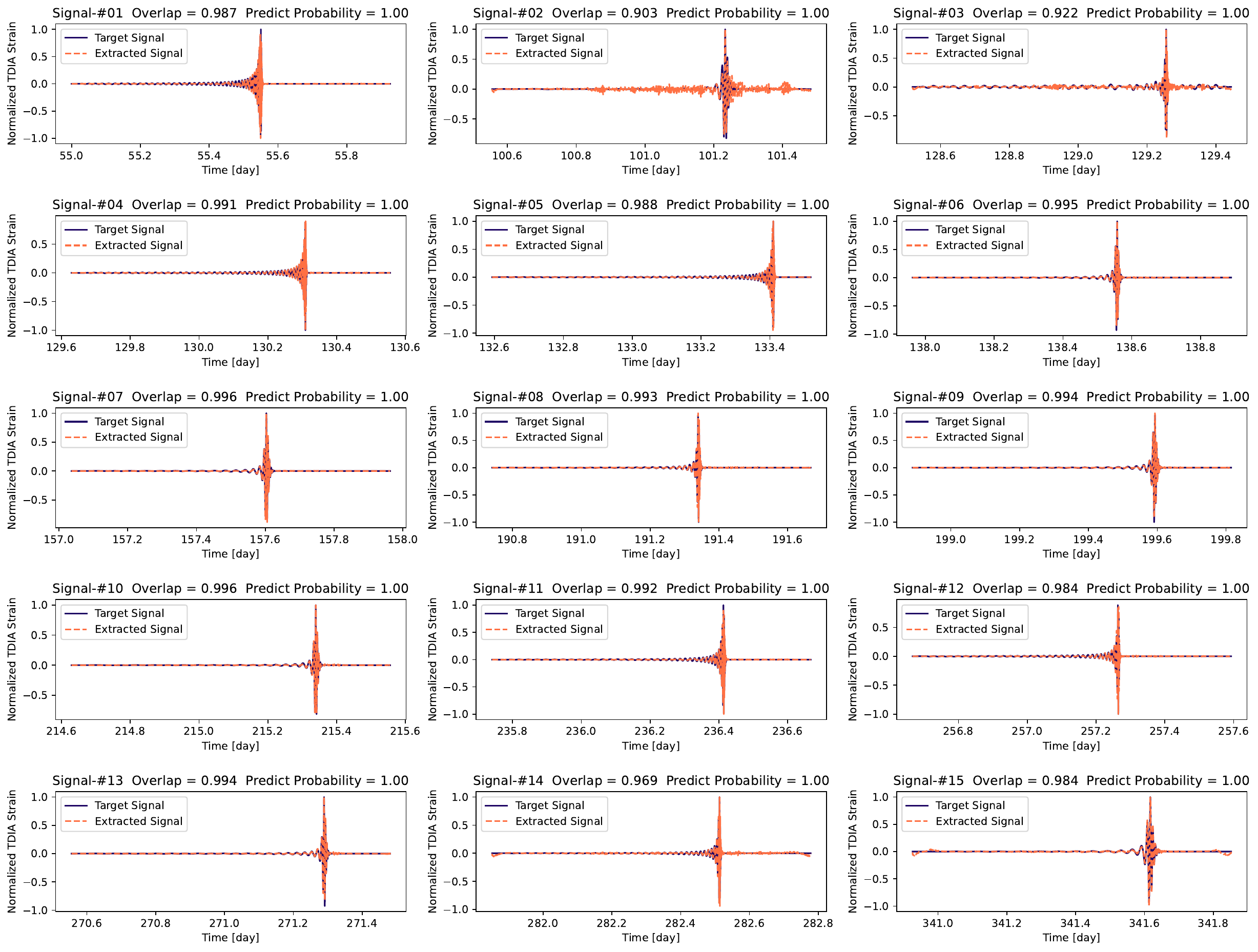}
    \caption{
        \label{fig:ldc2a-denoise} \textbf{Results of denoising and detection for the LDC 2a dataset using our model.} Each panel represents one of the 15 signals, with the number of the MBHB signal from the LDC2a dataset displayed at the top, along with the computed overlap between the denoised waveform generated by our model and the target template. Additionally, the predicted detection probability is provided. All 15 signals are effectively denoised, with most exhibiting high overlaps and detection probabilities equal to 1.
    }
\end{figure}

In the subsequent subsection, we provide an empirical assessment of our model employing the LDC2a dataset. As depicted in Figure \ref{fig:ldc2a-denoise}, the model proficiently extracts all 15 signals, with an overlap exceeding 0.9 for each, and notably, 13 of these signals exhibit an overlap above 0.96. Concurrently, the model demonstrates a detection probability equal to 1. The signal \#2 has a lower overlap because it is totally buried in the confusion noise.

When compared with MFCNN \cite{ruan_rapid_2023}, our approach not only ensures the detection of all 15 signals but also generates denoised waveforms with an overlap greater than 0.9, underlining its efficacy and precision.

On the other hand, in order to compare with the traditional method \cite{cornish_black_2020}, we run the MCMC test on the LDC2a dataset using the open source code from their repository\footnote{\url{https://github.com/eXtremeGravityInstitute/LISA-Massive-Black-Hole}} and compare the signal reconstructed by the best-fit parameters and the signal extracted by our neural network, the result is presented in Table \ref{tab:compare-mbhb}. The test results show that our neural network has comparable waveform extraction accuracy to the traditional method.

\begin{table}[htbp]
    \centering
    \caption{The table presents the overlap between the waveform reconstructed by the MCMC method and the template compared with the overlap between the waveform reconstructed by our neural network and the template. }
    \begin{tabular}{c|cccccccc}
        \hline
        \toprule
        Signal number & 1     & 2     & 3     & 4     & 5     & 6     & 7     & 8     \\
        \midrule
        MCMC          & 0.916 & 0.969 & 0.938 & 0.923 & 0.918 & 0.942 & 0.946 & 0.932 \\
        Denoised      & 0.987 & 0.903 & 0.922 & 0.991 & 0.988 & 0.995 & 0.996 & 0.993 \\
        \hline
        \midrule
        Signal number & 9     & 10    & 11    & 12    & 13    & 14    & 15            \\
        \midrule
        MCMC          & 0.957 & 0.951 & 0.925 & 0.938 & 0.944 & 0.940 & 0.958         \\
        Denoised      & 0.994 & 0.996 & 0.992 & 0.984 & 0.994 & 0.969 & 0.984         \\
        \bottomrule
        \hline
    \end{tabular}%
    \label{tab:compare-mbhb}%
\end{table}%

\subsection{Model inference speed}
The major advantage of deep neural networks compared with the traditional method is the speed. Table. \ref{tab:comput-cost} summarizes the computational cost of traditional data analysis methods based on technical notes submitted by various research groups in MLDC Round 1, Round 1B, and Round 3 as well as related papers. We couldn't find the MLDC Round 2 and 4 technical notes. We can see that traditional approaches for searching for GW signals inside 1 or 2-year MLDC and LDC datasets typically take a few hours. The computational cost of our method is presented in Table. \ref{tab:our-comput-cost}. While our model can evaluate 474 days of data in 1.9 seconds, it takes 2.5 minutes to evaluate the entire test dataset, which contains 101250 data samples (79 iterations with a batch size of 256).
The network could handle a batch of samples in parallel in a single computing iteration (see the 4th and 5th row of Table. \ref{tab:our-comput-cost}) (here batch means the number of data samples input to the network per computing iteration), but this does add up the signal duration time and it only means the model can process independent data segments in parallel with more computing resources.

\begin{table}[htbp]
    \centering
    \caption{Computational cost of traditional method on MLDC and LDC datasets}
    \small
    \begin{tabular}{lcccc}
        \toprule
        Challenge        & Group           & Method             & Computational Speed                        & Ref                                                    \\
        \midrule
        MLDC 1.1         & Ames            & Grid search        & 12 core-hours                              & \cite{thompson_mock_nodate}                            \\
        MLDC 1.1.1       & GLIG            & MCMC               & Each likelihood takes 31 seconds           & \cite{global_lisa_inference_group_report_nodate}       \\
        MLDC 1.1.2/1.1.3 & UTB             & Tomographic method & 18 core-hours                              & \cite{nayak_tomographic_nodate}                        \\
        MLDC 1.1.1a      & Cornish-Crowder & BAM                & $\sim 6$ core-hours                        & \cite{crowder_lisa_2006}                               \\
        MLDC 1.1.1b      & Cornish-Crowder & BAM                & $\sim 17$ core-hours                       & \cite{crowder_lisa_2006}                               \\
        MLDC 1.1.1c      & Cornish-Crowder & BAM                & $\sim 3$ core-hours                        & \cite{crowder_lisa_2006}                               \\
        MLDC 1.2.1       & Cornish-Porter  & MCMC               & $\sim 24$ core-hours                       & \cite{cornish_techincal_nodate,cornish_catching_2007}. \\
        MLDC 1.2.2       & Cornish-Porter  & MCMC               & $\sim 6$ core-hours                        & \cite{cornish_techincal_nodate,cornish_catching_2007}. \\
        MLDC 2.2         & MTAEI           & MCMC               & $\sim 3$ core-hours                        & \cite{cornish_search_2007}                             \\
        MLDC 3.2         & GSFC            & Xspec              & $\sim 36$ core-hours                       & \cite{arnaud_technical_nodate}                         \\
        MLDC 3.4         & CAM             & MULTINEST          & $\sim 2$ core-hours for 32768 seconds data & \cite{feroz_multinest_2009,feroz_technical_nodate}     \\
        LDC 1-1          & Cornish-Shuman  & MCMC               & few core-hours                             & \cite{cornish_black_2020}                              \\
        \bottomrule
    \end{tabular}%
    \label{tab:comput-cost}%
\end{table}%

\begin{table}[htbp]
    \begin{center}
        \caption{Computational cost of our method}\label{tab:our-comput-cost}%
        \begin{tabular}{@{}lcc@{}}
            \toprule
            Setting                                       & Our work                                                      \\
            \midrule
            Input data length [number of sampling points] & 16000                                                         \\
            Input data time duration                      & 44.4 hours                                                    \\
            Number of samples in a batch                  & 256                                                           \\
            Sampling points in a batch                    & $256 \times 16000 \simeq 4 \times 10^6$                       \\
            Data time duration in a batch                 & $44.4 \,\mathrm{hours} \times 256 \simeq 474\, \mathrm{days}$ \\
            Computational speed for a batch               & 1.9 seconds                                                   \\
            Computational speed for our test dataset      & 2.5 minutes for 79 iterations (batches)                       \\

            \bottomrule
        \end{tabular}
    \end{center}
\end{table}

\subsection{Model generalization behavior}
\begin{figure}[htbp]
    \centering
    \begin{subfigure}[t]{0.99\textwidth}{
        \includegraphics[width=\textwidth,valign=t]{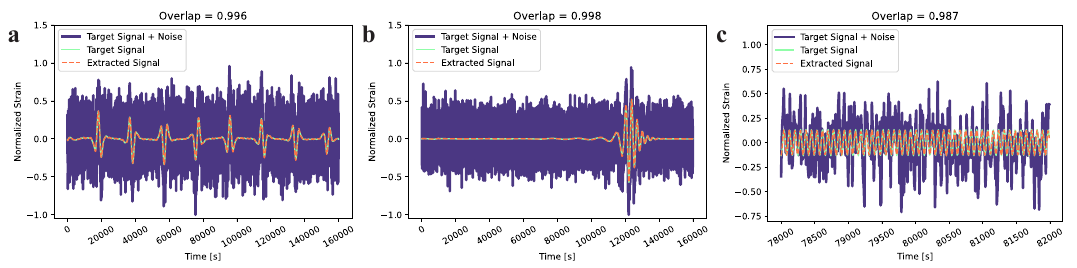}}
    \end{subfigure}
    \caption{
        \textbf{Showcases of the generalization behavior of our method.} \textbf{a}, EMRI signal with different semi-latus. \textbf{b}, MBHB signal with eccentricity. \textbf{c}, BWD signal with the mass transfer.
        The extracted waveform is compared with whitened templates. Only the middle part of the BWD waveform is presented to show the details of the waveform. The overlap between extracted data and waveform templates is shown on the top.
    }
    \label{fig:generation}
\end{figure}

Next, we evaluate the generalization ability of the model. Results are shown in Fig. \ref{fig:generation}.
The parameter space of our EMRI training dataset is only 4-dimensional with a fixed initial semi-latus rectum $p_0=20M$. Fig. \ref{fig:generation}a shows the result of testing our model using a signal with $p_0=30M$, which indicates a strong generalization capability beyond the training parameter space.
The training dataset of MBHB only contains GW signals from spin-aligned MBHB systems with quasi-circular orbits, without considering the case of orbital eccentricity.
Here we generate a GW signal with initial orbital eccentricity $e_0 = 0.5$ using the \texttt{SEOBNRE} model \cite{Liu_2022}.
In Fig. \ref{fig:generation}b, we can see the extracted effect, which demonstrates that our model has good generalizability. To test the generalization ability of the BWD model, our test signal is generated using an evolving BWD system that considers mass transfer, tidal forces, and gravitational radiation effects \cite{Kremer_2017}. The extracted result is shown in Fig. \ref{fig:generation}c. The output detection statistics labeled as detection probabilities of these three test signals are all equal to $1$.
The final test case is the SGWB model, here we consider broken power law signals following equation \eqref{eq:sgwb_bpl} with parameters $\alpha=-11.18$, $n_{t1}=2/3$, $n_{t2}=-1/3$, and $f_T=0.002$. This spectral shape might arise from the combination of two physically distinct sources. The classification test obtained ${\rm AUC}=0.99997$.

Then we evaluate our models' generalization ability to weaker signals. For EMRI, MBHB, and BWD models, we test them using data with lower SNR, for the SGWB model, we use test data with smaller amplitude. The histogram of overlap between extracted signal and the template is shown in Fig. \ref{fig:denoise}, at the same time the ROC curve of signal detection results is shown in Fig. \ref{fig:detection}. Throughout this series of tests, we have demonstrated the generalizability of our models in a wide variety of scenarios.

From an astrophysical perspective, first, LISA will observe MBHBs with very high SNR, typically bigger than 100, out to very high redshift \cite{bambi_space-based_2021}.
Then for EMRI signals, due to its physical complexity, the detection SNR threshold is $\sim 30$ \cite{bambi_space-based_2021}. Next, for the LISA verification binaries almost half of them reach a $ {\rm SNR} \ge 30$ \cite{10.1093/mnras/sty1545}. Finally, the upper limit of the SGWB is $\Omega_{GW}(25{\rm Hz}) \le 3.4 \times 10^{-9}$ for the case of $n_t=2/3$, corresponding to $\alpha = -11.74$ \cite{PhysRevD.104.022004}. In summary, the generalization ability of the model shows the potential of our model to be applied in practical situations.

\section{Discussion}
\label{sec:discussion}
With the results presented above, we show the efficacy of our Transformer-based deep neural network for space-based GW signal extraction and detection of multiple sources. Our method is reusable for either simulated or future observational data and gives an almost real-time analysis with low computational cost.

One potential limitation of our method is generalization behavior. We show the generalization performance for multiple signals outside the training dataset, those cases are still simpler than realistic astrophysical conditions. For example, the time delay interferometry technique is normally required to suppress the laser frequency noise in the space-based GW data analysis, which introduces further complexity in the detector response to the GW signal. Therefore, from a deep learning perspective, the patterns in the time domain might be different and a re-training of our neural network is required.

In this paper, we present a pioneering proof-of-principle study that utilizes DNNs for the efficient detection and extraction of space-based gravitational wave signals. It's essential to clarify that our neural network model is not aimed at replace conventional matched filtering techniques. Instead, it seeks to offer an efficient way of processing the potentially large amount of data from space-based detectors, thereby facilitating more automated and real-time gravitational wave data analysis.

Our model has demonstrated promising results in analyzing MBHB signals, even in datasets with lower signal-to-noise ratios than LDC datasets. 
Nonetheless, our approach still faces limitations when addressing long-lasting signals, such as EMRIs and BWDs, which accumulate signal-to-noise ratios over time spans on the order of years. Given the current computational resources, analyzing these signals in a singular pass poses a considerable challenge

Despite the challenges faced, this research stands as a stepping stone in the domain of GW data analysis. By laying the foundation for future endeavors, we are aiming at the continual development and optimization of deep learning techniques in this field. As computational resources and technology advance, our model holds the potential for adaptation and evolution to effectively tackle these challenges.

\section{Methods}
\label{sec:method}
\subsection{GW sources in space-based detection}
As mentioned in the section \ref{sec1}, space-based  GW detectors are being developed to detect GW signals at frequencies of $[10^{-4},0.1] {\rm Hz}$. The main GW sources in this frequency band are EMRI, MBHB, BWD, and SGWB. Next, we will describe the details of the signals that come from each GW source.

\subsubsection*{EMRI}
The MBHs in the centers of galaxies are typically surrounded by clusters of stars. These stars eventually evolved into COs, which will be black holes, neutron stars, or white dwarfs. Some of those COs can get captured onto orbits bound to the central MBH and then gradually inspiral into the MBH via emission of GWs. Typically the ratio of the mass of the CO that is falling into the MBH to the mass of the MBH is $\sim 10^{-5}$, so these events are called EMRIs.

EMRI signal waveforms are characterized by the complex time domain strain $h(t)$, in the source frame $h(t)$ is given by \cite{katz_fastemriwaveforms_2021}:
\begin{equation}
    h(t)=\frac{\mu}{d_L}\sum_{lmkn}A_{lmkn}(t)S_{lmkn}(t,\theta)e^{im\phi}e^{-i\Phi_{mkn}(t)},
\end{equation}
where $\mu$ is the mass of the small black hole, $t$ is the time of arrival of the GW, $\theta$ is the source-frame polar angle, $\phi$ is the source-frame azimuthal angle, $d_L$ is the luminosity distance, and \{l; m; k; n\} are the indices describing harmonic modes. The indices l, m, k, and n label the orbital angular momentum, azimuthal, polar, and radial modes, respectively.
$\Phi_{mkn}=m\Phi_\varphi+k\Phi_\theta+n\Phi_r$ is the summation of phases for each mode. $A_{lmkn}$ is the amplitude of GW.
$S_{lmkn}(t,\theta)$ is spin-weighted spheroidal harmonic function.

In the detector frame, the EMRI signal waveform is determined by 17 parameters: $\{M,\mu, a, {\vec{a}}_2, p_0, e_0, x_{I,0}, d_L,\theta_S, \phi_S, \theta_K, \phi_K, \Phi_{\varphi,0}, \Phi_{\theta,0}, \Phi_{r,0} \}$.
$M$ is the mass of the MBH, $a$ is the dimensionless spin of the MBH, $\theta_S$, and $\phi_S$ are the polar and azimuthal sky location angles.
$\theta_K$ and $\phi_K$ are the azimuthal and polar angles describing the orientation of the spin angular momentum vector of the MBH.
${\vec{a}}_2$ is the spin vector of the CO, which doesn't considered in the waveform model.
$p$ is semi-latus rectum, $e$ is eccentricity,  $I$ is orbit’s inclination angle from the equatorial plane, and $x_I\equiv \cos I$.

\subsubsection*{MBHB}
Most galaxies appear to host black holes at their centers. Galaxies and MBHBs coevolved during the evolution of the Universe.
So the observation of GWs from the MBHB system can improve our understanding of important astronomical phenomena such as the formation of the MBH and the merging of galaxies \cite{bambi_space-based_2021}.

In this paper we just consider the GW from spin-aligned MBHB system, which characterized by $\{M_{tot},q,s^z_1,s^z_2\}$, where  $M_{tot} = m_1 + m_2$, $m_1$ and $m_2$ are the mass of two black holes respectively.
$q=m_2/m_1$ ($m_1 > m_2$) is the mass ratio. $s^z_1$ and $s^z_2$ are spin parameters of two black holes, and $z$ represent the direction of orbital angular momentum.

\subsubsection*{BWD}
The Milky Way contains a large population of compact binaries, most of which are BWD with a period of $\sim 1 hour$.
This is right in the LISA's sensitive frequency band. The signal of BWD in the source frame is quite simple:
\begin{equation}
    \begin{aligned}
        \label{bwd_wf}
        h_{+}(t)    & =\mathcal{A}(1+\cos^2\iota)\cos\Phi(t), \\
        h_\times(t) & = 2\mathcal{A}\sin\iota\sin\Phi(t),     \\
        \Phi(t)     & =\phi_0+2\pi f t +\pi \dot{f} t^2.
    \end{aligned}
\end{equation}
$\mathcal{A}$ is the overall amplitude, $\phi_0$ is the initial phase at the start of the observation, $\iota$ is the inclination of the BWD orbit to the line of sight from the origin of the Solar System Barycentric (SSB) frame. The intrinsic parameter is the frequency of the signal $f$ and its derivative $\dot{f}$.

Frequency evolves slowly and some binaries will be chirping to higher frequency due to the decay of the orbit through the emission of GWs, but other binaries will be moving to lower frequency as a result of mass transfer between the binary components driving an increase in the orbital separation \cite{bambi_space-based_2021}.

\subsubsection*{SGWB}
There are many resolvable sources, but there is also a large number of events that cannot be resolved individually, resulting in SGWB.
Astrophysical background components are guaranteed in the LISA band, originating from unresolved Galactic Binaries (GB) and stellar-originated black hole mergers.
SGWBs that are Gaussian, isotropic, and stationary can be fully described by their spectrum \cite{caprini_reconstructing_2019}:
\begin{equation}
    \Omega_{GW}(f)=\frac{f}{\rho_c}\frac{{\rm d}\, \rho_{GW}}{{\rm d}\,f},
\end{equation}
where $\rho_{GW}$ is the energy density of gravitational radiation contained in the frequency range $f$ to $f +df$, $\rho_c = \frac{3H_0^2c^2}{8\pi G}$ is the critical density of the universe, where $c$ is the speed of light, and $G$ is Newton’s constant, $H_0=67.9{\rm km \ s^{-1\ } Mpc^{-1}}$ is the Hubble constant.

We intend to follow the simplified assumption that the signal can be well described by a power law, defined as amplitude and slope, as most studies have done previously. Then the signal is described by \cite{flauger_improved_2021}:
\begin{equation}
    \label{eq:sgwb_pl}
    h^2\Omega_{GW}(f)=10^\alpha\left(\frac{f}{f_*}\right)^{n_t},
\end{equation}
where $h$ is dimensionless Hubble constant, $f_*$ is pivot frequency, $\alpha$ characterize its amplitude at $f_*$ and $n_t$ is the slope of spectrum.

Another formalism used to test our model's generalization ability is broken power law which is defined:
\begin{equation}
    \label{eq:sgwb_bpl}
    h^2\Omega_{GW}(f)=10^\alpha \left[ H(f_T-f) \left(\frac{f}{f_*}\right)^{n_{t1}} + H(f-f_T)\left(\frac{f_T}{f_*}\right)^{n_{t1}} \left(\frac{f}{f_T}\right)^{n_{t2}}\right],
\end{equation}
where $n_{t1}$ and $n_{t2}$ are the slopes of two spectrum segments, $H(f)$ is Heaviside step function.

\subsection{Data Curation}
\label{sec:data}
First, we simulate the noise data from the LISA sensitivity curve:
\begin{equation}
    S_n(f) = \frac{1}{L^2\mathcal{R}(f)} \left(P_{\rm OMS} + 2(1+\cos^2(f/f_*))\frac{P_{\rm acc}}{(2\pi f)^4}\right),
\end{equation}
where
\begin{align}
    P_{\rm OMS} & = (1.5 \times 10^{-11}\, {\rm m})^2 \left(1 + \left(\frac{2\,{\rm mHz}}{f}\right)^4\right)\,{\rm Hz^{-1}}
\end{align}
and
\begin{align}
    P_{\rm acc} & = (3 \times 10^{-15}\, {\rm m})^2 \left(1 + \left(\frac{0.4\,{\rm mHz}}{f}\right)^2\right)\left(1 + \left(\frac{f}{8\,{\rm mHz}}\right)^4\right)\,{\rm Hz^{-1}}
\end{align}
are the optical metrology noise and acceleration noise respectively. $L=2.5 \times 10^{9}\, {\rm m}$, $f_*=19.09\, {\rm mHz}$, $\mathcal{R}(f)$ is the transfer function. The full expression for $\mathcal{R}(f)$ used here is computed numerically \cite{robson_construction_2018}.
Next we specify the generation of each dataset.

We use the augmented analytic kludge (AAK) \cite{chua_improved_2015,chua_fast_2017, katz_fastemriwaveforms_2021} model to generate the EMRI signal.
It is because the AAK model combines the accuracy of the numerical kludge (NK) model and the computational speed of the analytic kludge (AK) model quite well.
Note that the parameter $\iota$ in the AAK model is a orbital parameter: $\cos \iota = L_z/\sqrt{L_z^2+Q}$, where $Q$ is Carter constant, $L_z$ is $z$  component of the specific angular momentum.
For simplicity, we just consider a small parameter space to generate the training data. Detailed parameters range is shown in Table \ref{tab:emri_par}, and the lower bound of $a$ and $e_0$ is limited by the \texttt{FastEMRIWaveform} toolkit we used \cite{michael_l_katz_2020_4005001}.

\begin{table}[htbp]
    \begin{center}
        \caption{Summary of parameter setups in EMRI signal simulation.}\label{tab:emri_par}%
        \begin{tabular}{@{}ccc@{}}
            \toprule
            Parameter   & Lower bound   & Upper bound   \\
            \midrule
            $M$         & $10^5M_\odot$ & $10^7M_\odot$ \\
            $a$         & $10^{-3}$     & $0.99$        \\
            $e_0$       & $10^{-3}$     & $0.5$         \\
            $\cos\iota$ & $-1$          & $1$           \\
            \bottomrule
        \end{tabular}
    \end{center}
\end{table}

For MBHB signal generation, we used \texttt{SEOBNRv4\_opt}, which is a version of the \texttt{SEOBNRv4} code \cite{bohe_improved_2017} with significant optimizations.
It could produce signals for a high spin, high mass ratio MBHB system.
We adopted the log-uniform distribution for the parameter $M_{tot}$ from Ref. \cite{katz_fully_2022}.
Detailed parameters range is shown in Table \ref{tab:mbhb_par}.

\begin{table}[htbp]
    \begin{center}
        \caption{Summary of parameter setups in MBHB signal simulation.}\label{tab:mbhb_par}%
        \begin{tabular}{@{}ccc@{}}
            \toprule
            Parameter & Lower bound   & Upper bound   \\
            \midrule
            $M_{tot}$ & $10^6M_\odot$ & $10^8M_\odot$ \\
            $q$       & $0.01$        & $1$           \\
            $s_1^z$   & $-0.99$       & $0.99$        \\
            $s_2^z$   & $-0.99$       & $0.99$        \\
            \bottomrule
        \end{tabular}
    \end{center}
\end{table}

For the BWD dataset, we generate the signal directly using equations \eqref{bwd_wf}.
We follow the parameter setting of the LISA Data Challenge (LDC) 1-4 dataset \cite{zhang_resolving_2021}, but we focus only on the intrinsic parameters $f$ and $\dot{f}$, see Table \ref{tab:bwd_par} for details.

\begin{table}[htbp]
    \begin{center}
        \caption{Summary of parameter setups in BWD signal simulation.}\label{tab:bwd_par}%
        \begin{tabular}{@{}ccc@{}}
            \toprule
            Parameter & $f$                & $\dot{f}$                                       \\
            \midrule
            range-1   & $[0.1,4]{\rm mHz}$ & $[-3\times10^{-17},6\times10^{-16}] {\rm Hz^2}$ \\
            range-2   & $[4,15]{\rm mHz}$  & $[-3\times10^{-15},4\times10^{-14}]{\rm Hz^2} $ \\
            \bottomrule
        \end{tabular}
    \end{center}
\end{table}

Upon generating the signal, we project it onto the LISA detector. For this work, being a proof-of-concept, we did not incorporate the time delay interferometry (TDI) technique. Instead, we employed the long-wavelength approximation, as described below:
\begin{equation}
    \begin{aligned}
        h_{I,II}(t) = \frac{\sqrt{3}}{2} \left(h_+(t)F^+_{I,II}(\theta_S,\phi_S,\psi_S) \right.
        \\\left. + h_\times(t)F^{\times}_{I,II}(\theta_S,\phi_S,\psi_S) \right),
    \end{aligned}
\end{equation}
where $F^+_{I,II}$ and $F^{\times}_{I,II}$ are the antenna pattern functions:
\begin{equation}
    \begin{aligned}
        F^+_{I}(\theta_S,\phi_S,\psi_S) = \frac{1}{2}(1+\cos^2\theta_S)(\cos 2\phi_S \cos 2\psi_S        \\- \cos\theta_S \sin 2 \phi_S \sin 2\psi_S),\\
        F^{\times}_{I}(\theta_S,\phi_S,\psi_S) = \frac{1}{2}(1+\cos^2\theta_S)(\cos 2\phi_S \sin 2\psi_S \\+ \cos\theta_S \sin 2 \phi_S \cos 2\psi_S),
    \end{aligned}
\end{equation}
and
\begin{equation}
    \begin{aligned}
        F^+_{II}(\theta_S,\phi_S,\psi_S) = F^{+}_{I}(\theta_S,\phi_S,\psi_S-\pi/4), \\
        F^{\times}_{II}(\theta_S,\phi_S,\psi_S) = F^{\times}_{I}(\theta_S,\phi_S,\psi_S-\pi/4),
    \end{aligned}
\end{equation}
where $\psi_S$ is the polarization angle.
Furthermore, this antenna pattern function varies with time as a result of the motion of the LISA detector.
In this paper, we set the sky position and the polarisation angle equal to zero for simplicity. Because of the length of our data, the Doppler shift effects could be ignored.

For those 3 datasets, we inject the signal to the noise with specific optimal SNR as:
\begin{align}
    \label{eq:snr}
    {\rm SNR} = \left(s\mid s\right)^{1/2},
\end{align}
where $s$ represent the signal template, the inner product $(h\mid s)$ is:
\begin{equation}
    (h \mid s) = 2\int_{f_{min}}^{f_{max}} \frac{\tilde{h}^*(f)\tilde{s}(f)+\tilde{h}(f)\tilde{s}^*(f)}{S_n(f)}\, df,
\end{equation}
where $f_{min}=3\times 10^{-5} {\rm Hz}$ and $f_{max}=0.05{\rm Hz}$. $\tilde{h}(f)$ and $\tilde{s}(f)$ are frequency domain signals and the superscript $*$ means complex conjugate, $S_n(f)$ is the noise PSD.
Here following the setting of the LDC dataset, we set the SNR of the training data to $50$.
Then the data was whitened and normalized to $[-1,1]$. During the whitening procedure, we applied the Tukey window with $\alpha=1/8$.

This inner product can also be used to measure how well the output of our model matches the signal waveform template, we calculate the overlap between them, which is defined as:
\begin{align}
    \label{eq:overlap}
    \mathcal{O}(h,s)=\left(\hat{h} \mid \hat{s}\right)
\end{align}
with
\begin{align}
    \hat{h}=h \left(h\mid h\right)^{-1/2},
\end{align}
where $h$ represent the model output and $s$ represent the template.

The SGWB dataset is generated in a very different way than several other datasets, We just need to simulate SGWB data based on its PSD which is defined by:
\begin{align}
    S_h(f) = \frac{3H_0}{4\pi^2}\frac{\Omega_{GW}(f)}{f^3}.
\end{align}
We choose $n_t=2/3$ $\alpha = -11.35$ and $f_*=10^{-3}{\rm Hz}$ (see equation \eqref{eq:sgwb_pl}) according to LDC configuration and Ref. \cite{flauger_improved_2021} , which represent SGWB formed by compact binary coalescences.
We could generate the SGWB signal by $S_h(f)$ directly.
We then combine the signal and noise and perform the whitening and normalization operations as described above.
Lastly, we present the pure signals with noise PSD and SGWB PSD in the Fourier domain and the data generated to train our model in Fig. \ref{fig:data}.

\subsection{Transformer}
Transformer\cite{vaswani2017attention} is a kind of deep neural network (DNN) proposed for machine translation, and it soon achieved superior performance in various tasks in natural language processing\cite{devlin2018bert} and computer vision\cite{dosovitskiy2020image}. Based entirely on attention, Transformer has a great ability to capture both long-range and short-range dependency in sequence data. In this section, we introduce the key structures of a general Transformer.

\subsubsection*{Self-Attention}
Self-Attention can be described as a function with an input vector Query and an output vector pair Key-Value. The Key-Value pair is a weighted sum, which returns the information of the Query with the corresponding Key. In the Transformer network block, all query vectors and key-values pairs have the same dimension $d$. Given a sequence of queries $Q \in \mathbb{R}^{l\times d}$ with length $l$, keys $K \in \mathbb{R}^{l\times d}$ and values $V \in \mathbb{R}^{l\times d}$, the Transformer compute scaled dot-product attention as:
\begin{align}
    {\rm Attention}{(Q,K,V)}={\rm softmax}(\frac{Q K^T}{\sqrt{d}})V.
\end{align}

\subsubsection*{Multi-Head Attention}
Instead of applying a single attention function with $d_{model}$-dimensional queries, keys, and values. The Transformer uses multi-head attention to combine information from different linear projections of original queries, keys, and values. 

If the Transformer has $H$ heads, the sequence of attention output $head_i$ is:
\begin{align}
    head_i={\rm Attention}{(Q_i,K_i,V_i)}={\rm softmax}(\frac{Q_i K_i^T}{\sqrt{d}})V_i,
\end{align}
where $Q_i=QW_i^Q$, $K_i=KW_i^K$, $V_i=VW_i^V$, are projected queries, keys and values, corresponding to head $i \in \{1,...,H\}$ with learning parameters $W_i^Q,W_i^K,W_i^V$, respectively.
Here,
\begin{align}
    \label{eq:att_w}
    A=\frac{1}{H}\sum_{i=1}^{H}{\rm softmax}(\frac{Q_i K_i^T}{\sqrt{d}})
\end{align}
is also called attention map. $A \in \mathbb{R}^{l\times l}$, in which each element $A_{qk}$ indicates how much attention token $q$ puts on token $k$.
With collection of all parameters $W^H \in \mathbb{R}^{Hd_{model}\times d}$, the multi-head attention (MHA) of these H heads can be written as:
\begin{align}
    {\rm MHA}{(Q,K,V)}={\rm Concat}(head_1,...,head_H)W^H.
\end{align}
The multi-head attention mechanism can be computed in parallel for each head, which leads to high efficiency.
Moreover, multi-head attention connects the information from different projection subspaces directly, helping the Transformer learn the long-term dependencies of the input sequence easier.

\subsubsection*{Feed Forward and Residual Connection}
In addition to attention layers, Transformer blocks have a fully connected feed-forward network that operates separately and identically on each position:
\begin{align}
    {\rm FFN}(H')={\rm ReLU}(H'{W}_1+b_1)W_2+b_2,
\end{align}
where $H'$ is the output of last layer, and $W_1, b_1, W_2, b_2$ are trainable parameters.
The dimension of input and output is equal to the model's dimension $d_{model}$, and the inner-layer dimension $d_{ffn}$ should be larger than $d_{model}$.
In a deeper Transformer model, a residual connection module is inserted followed by a Layer Normalization Module. The output of the Transformer block can be written as:
\begin{align}
    H'={\rm LayerNorm}({\rm Attention}(X)+X), \\
    H={\rm LayerNorm}({\rm FFN}(H')+H').
\end{align}

\subsection{Network structure}
Let the T-length observation $x\in \mathbb R^T$ be a time-domain signal we receive.
$x$ is a mixture of a target GW signal $s$ and noise $n$ as $x=s+n$, where the noise is from the environment and instruments. Our goal is to recover $s$ from $x$. The recovered signal $\hat s \in \mathbb{R}^T$ can be written as:
\begin{align}
    \hat s = {\rm Dec}({\rm Enc}(x)\otimes \mathcal{M}).
\end{align}
The decoder reconstructs the signal with encoded input $x$ element-wise multiplication by the mask $\mathcal{M}$ predicted by the masking net. After recovering the signals, we add a multi-layer linear perception to detect whether it is a GW signal or a pure noise.
\subsubsection*{Encoder and Decoder}
We use a CNN as an encoder because it can extract local features from a long time-domain sequence, which could compress information. With time-domain input $x\in \mathbb R^T$, the encoded ${\rm Enc}(x) \in \mathbb R^{L\times T'}$. Since the output estimated signal has the same length as input $x \in \mathbb R^T$, the decoder for reconstruction uses a transposed convolution layer.

\subsubsection*{Masking Extraction Net}
The masking network is built by following a basic structure in SepFormer\cite{subakan2021attention}.
We employ two Transformer blocks the STTB (Short-Time Transformer Block) and the LTTB (Long-Time Transformer Block) in the masking net. The masking network is fed by the encoded input. We split the input signal into chunks and concatenated them to be a tensor $g \in \mathbb R^{L\times C\times N}$, where $C$ is the length of each chunk and $N$ is the number of chunks.

The tensor $g$ is processed by Transformer blocks.
The STTB computes the multi-head attention in each chunk respectively, which catches the short-time dependency in the chunk.
Then the LTTB focuses on another dimension of tensor $g$, modeling the long-time dependency by the attention across chunks.
The output from Transformer blocks passes through a PReLU non-linearity layer and a 2-D convolution layer for matching the output size of the decoder.
Then a two-path convolution with different linear functions is used to get the mask.
\subsubsection*{Multi-Layer Perception}
The extracted signal recovered by the decoder feeds to the MLP for detection. We use two fully connected layers to classify GW signals and noise. The first linear layer has 512 dimensions, and the second linear layer outputs the vector to a probability of a true GW signal.

\subsection{Loss function}
Our loss function combines both the extraction loss and the detection loss. The extraction loss is based on the scale-invariant signal-to-distortion ratio\cite{vincent2006performance} in audio enhancement, which is defined as:
\begin{equation}
    \begin{aligned}
        s_{target}       & =\frac{\hat{x}^Tx}{\Vert x \Vert^2},               \\
        \mathcal{L}_{EX} & =10\log_{10}\frac{s_{target}}{\hat{x}-s_{target}},
    \end{aligned}
\end{equation}
where $\hat{x}$ is the estimated output and the $x$ is the target.

The detection loss is the binary cross-entropy. Suppose that the data set has $N$ samples with label $y$, and the $\hat{y}$ is the predicted probability of the sample. The BCE loss is defined as:
\begin{align}
    \mathcal{L}_{DE}=-\frac{1}{N}\sum_{i=1}^{N}y_i\log(\hat{y_i})+(1-y_i)\log(1-\hat{y_i}).
\end{align}

Therefore, the total loss of the deep neural network is:
\begin{align}
    \mathcal{L}=\mathcal{L}_{EX}+\mathcal{L}_{DE}.
\end{align}

\subsection{Implementation details}
Our extraction network repeats both STTB and LTTB twice ($M=2$), with 4 parallel attention heads and a 512-dimensional feed-forward layer in each block. We set the length of split chunks $C=25$. In the training stage, the model is trained for 100 epochs. We set initial learning rate as ${lr}_{max}=1e^{-3}$. After 35 epochs, the learning rate is annealed by halving if the validation performance does not improve for two generations.
Adam\cite{kingma2014adam} is used as the optimizer with $\beta_1=0.9,\beta_2=0.98$. The network is trained on a single NVIDIA V100 GPU. All of our code is implemented in \texttt{Python} within the \texttt{SpeechBrain} \cite{ravanelli2021speechbrain} toolkit.

\section{Data avalibility}
\label{sec:data-aval}
The datasets used in this study are generated by the custom code, which is provided in the repository mentioned in the 
\nameref{sec:code-aval}
section. To reproduce the datasets, please follow the instructions provided in our repository's documentation.

\section{Code availability}
\label{sec:code-aval}
The \texttt{PyCBC}, \texttt{FastEMRIWaveform}, and \texttt{SpeechBrain} codes used in this study are publicly available.
The custom code developed for this research can be accessed on GitHub at the following repository: \url{https://github.com/AI-HPC-Research-Team/space_signal_detection_1}. The code is distributed under the MIT License.

\bibliographystyle{apsrev4-1}
\bibliography{references}

\section{Acknowledgements}
The research was supported by the Peng Cheng Laboratory and by Peng Cheng Laboratory Cloud-Brain.
This work was also supported in part by the National Key Research and Development Program of China Grant No. 2021YFC2203001 and in part by the NSFC (No.~11920101003 and No.~12021003). Z.C was supported by ``The Interdisciplinary Research Funds of Beijing Normal University" and CAS Project for Young Scientists in Basic Research YSBR-006.

\section{Author contributions}
Z.R obtained the major funding and conceived this research. Z.C also support the funding and supervised the astrophysical science analysis in this research. T.Z and R.L designed the experiments and performed analyses. T.Z performed data generation and trained the network. R.L developed the detailed method of deep learning and implemented the network. T.Z and R.L wrote the manuscript with input from Z.R and Z.C. H.W assisted in designing the data processing software.

\section{Competing interests}
The authors declare no competing interests.

\end{document}